\documentstyle[sprocl,epsfig]{article}

\def\Journal#1#2#3#4{{#1} {\bf #2}, #3 (#4)}

\def\PLB{{\em Phys. Lett.}  B}

\def\PRD{{\em Phys. Rev.} D}
\def\ZPC{{\em Z. Phys.} C}
\def\be{\begin{equation}}
\def\ee{\end{equation}}
\def\bea{\begin{eqnarray}}
\def\eea{\end{eqnarray}}
\begin{document}

\begin{flushright}
PRA--HEP/97--10
\end{flushright}
\title{VIRTUAL PHOTON STRUCTURE FROM JET PRODUCTION}

\author{ J. CH\'{Y}LA
\footnote{Talk presented at PHOTON '97 Conference in Egmond aan Zee}
, J. CVACH}

\address{Institute of Physics of the Academy of Sciences,\\
 Na Slovance 2, Prague 8, 18040, Czech Republic}

\maketitle\abstracts{Some aspects of
extracting the information on the structure of virtual photons from
jet production in ep and e$^+$e$^-$ collisions are discussed.}

\section{Virtual particles - why, where and how?}
Measuring the dependence of the structure of the photon at short
distance on its virtuality 
provides new information on strong interaction dynamics.
The results of first attempts in this direction have recently
been reported\cite{data}.
Presently the structure of virtual photons can
be investigated via the deep inelastic scattering or jet production
in $\gamma\gamma$ collisions at LEP, or in jet production in $\gamma$p
collisions at HERA. In jet studies the internal
structure of the photon manifests itself through the resolved photon
contribution to hard $\gamma$p or $\gamma\gamma$ collisions.
Measuring the deviation of jet cross--sections from expectations
of a structureless photon provides information on
the distribution functions $f_{i/\gamma}(x,M^2,P^2)$ of partons
inside the photon of virtuality $P^2$, probed at the distance $1/M$.
In Fig. 1a we compare
the integrated cross--sections for the DIS and dijet
production
\begin{displaymath}
\sigma^{\mathrm DIS}(M^2) \equiv \int_{M^2}{\mathrm d}Q^2\int
{\mathrm d}x\frac{{\mathrm
d}\sigma(x,Q^2)} {{\mathrm d}x{\mathrm d}Q^2},\;\;
\sigma^{\mathrm jets}(M^2)\equiv \int_{M}{\mathrm d}p_T
\int{\mathrm d}\eta
\frac{{\mathrm d}\sigma(\eta,p_T)}{ {\mathrm
d}\eta{\mathrm d}p_T}
\end{displaymath}
in e$^+$e$^-$ collisions at LEP 2 for photons with 
$P^2<1$ GeV$^2$. The hard scale is 
identified with the standard $Q$ in DIS and $p_T$ in jet production.
The curves in Fig. 1a show that up to moderate scales jet
production is more effective than DIS, while for $M^2$ above 
40 GeV$^2$ DIS has larger cross--section. Although both 
cross--sections decrease with increasing $P^2$, their ratio 
is essentially independent of it. To obtain reasonable statistics
of a few thousands of events it will be necessary to use jets
with transverse momenta down to 3-4 GeV.
In \cite{my} we discussed the possibilities
offered by the HERA upgrade for the measurement of basic
features of the virtual photon structure via the jet production.
Here we elaborate on two important aspects of this problem.
\begin{figure}
\begin{center}
\epsfig{file=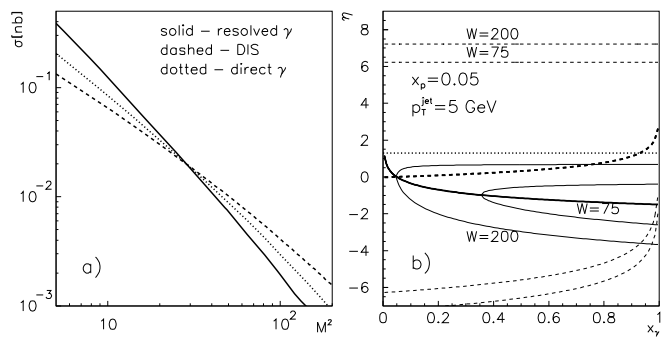,height=4.5cm}
\end{center}
\caption{Integrated cross--sections of $\gamma$ with $Q^2\le$ 1
GeV$^2$ for the DIS and jet production in e$^+$e$^-$ collisions
at LEP 2, using the GRV structure
functions for the photon and disregarding any detector efficiencies
(a), kinematical
ranges of the pseudorapidity of jets (solid lines) and beam
remnants at HERA (dashed) (b). In (b) the thick solid and dashed lines
indicate the centers of the respective intervals and the dotted one the
end of the H1 detector.}
\end{figure}

\section{WW approximation and the structure of the virtual photon}
The measured electroproduction cross--sections at the CMS energy
$\sqrt{S}$ are related
to those of the photoproduction via the WW approximation as
\begin{displaymath}
{\mathrm d}\sigma(e+p\rightarrow e'+X;S)=
\int\int {\mathrm d}y{\mathrm d}P^2
f_{\gamma/e}(y,P^2){\mathrm d}
\sigma(\gamma+p\rightarrow X;yS,P^2),
\end{displaymath}
where $f_{\gamma/e}(y,P^2)$ is the
unintegrated distribution function of virtual photons
inside an electron. This relation holds for $P^2$ much
smaller than the scale of the $\gamma+p\rightarrow X$
process. If this latter process is characterized by the hard scale $Q$,
its cross--section can in turn be expressed in terms of the
convolution of
integrated parton distribution functions (pdf)
$f_{i/\gamma}(x_{\gamma},Q^2,P^2)$ of the photon
with those of the proton, $f_{j/p}(x_p,Q^2)$,
and partonic hard scattering cross--section
$\sigma_{ij}$.
Because virtual photons can be transverse as well as longitudinal,
there are
two independent fluxes. Neglecting subdominant terms we have
\bea f^{T}_{\gamma/e}(x,P^2) & = &
\frac{\alpha}{2\pi}\frac{1+(1-x)^2}{x}\frac{1}{P^2} \equiv
\frac{F^T(x)}{P^2},
\label{T} \nonumber\\
f^{L}_{\gamma/e}(x,P^2) & = &
\frac{\alpha}{2\pi}\frac{2(1-x)}{x}\frac{1}{P^2} \equiv
\frac{F^L(x)}{P^2}.
\nonumber
\label{L}
\eea
As $P^2\rightarrow 0$, $f_{i/\gamma}^{T,L}$ behave as
\bea
f_{i/\gamma}^{T}(x,Q^2,P^2)& = & f_{i/\gamma}^{\mathrm real}(x,Q^2)+
(P^2/\mu^2_T)h_i^T(x,Q^2) +O(P^4), \label{apT}\nonumber \\
f_{i/\gamma}^{L}(x,Q^2,P^2)& = & (P^2/\mu^2_L)h_i^L(x,Q^2)+O(P^4),
\label{apL}
\nonumber
\eea
where $\mu_{T,L}$ are some parameters. In most of existing 
phenomenological 
analyses 
\footnote{The relevance of the longitudinal
part $f^L_{\gamma/e}$ has recently
been been discussed in \cite{Schuler}.}
of the photon structure its longitudinal componennt
$f^L$ has been neglected.
This is, however, consistent only if we at the same time
neglect also the dependence of
$f^T_{i/\gamma}(x,Q^2,P^2)$ on the virtuality $P^2$! Indeed, as
\bea
f_{\gamma/e}^T(P^2)\otimes f_{i/\gamma}^T(Q^2,P^2)& = &
(1/P^2)F^T\otimes f_{i/\gamma}^{\mathrm real}(Q^2)+
(1/\mu_T^2)F^T\otimes h_i^T(Q^2), \label{TT}\nonumber\\
f_{\gamma/e}^L(P^2)\otimes f_{i/\gamma}^L(Q^2,P^2)& = &
(1/\mu_L^2)F^L\otimes h_i^L(Q^2), \label{LL} \nonumber
\eea
$f^L_{\gamma/e}$ contributes terms of the same order as the subleading,
$O(P^2)$, term in $f^T_{\gamma/e}$! Thus to investigate $P^2$ dependence
of photon structure, we \underline{must}
 take into account also its longitudinal
component. Details will be discussed in \cite{my2}.
The $P^2$ dependence of the photon pdf has recently been addressed in
a number of papers \cite{BS,DG,SS,GRS}. For instance, in
\cite{DG} the virtuality dependence of $f^T_{\gamma/e}$ is assumed to
be given by a simple multiplicative suppression factor $L$
\begin{displaymath}
f_{q/\gamma}^T(x,Q^2,P^2)=L(Q^2,P^2,P_c^2)
f^{\mathrm real}_{q/\gamma}(x,Q^2),
\;L\equiv\frac{
\ln\left((Q^2+P_c^2)/(P^2+P_c^2)\right)}{
\ln\left((Q^2+P_c^2)/P_c^2\right)}.
\end{displaymath} 
As for $P^2\rightarrow 0$,
$L\approx 1-(P^2/P_c^2)/\ln(Q^2/P_c^2)$,
the parameter $P_c$ plays the role of $\mu_T$. For the
gluon the suppression factor is taken as $L^2$. This simple ansatz,
incorporated in HERWIG event generator, was used in \cite{my}.

\section{Jets vs. the underlying event in ep collisions at HERA}
Also at HERA the studies of resolved photon processes 
 with $P^2$ above, say,
$0.5$ GeV$^2$, will require using jets
with $E_T$ down to about 5 GeV$^2$. The properties of such low $E_T$ jets can, however,
be significantly distorted by the presence of soft particles produced
in the {\em soft underlying event} (sue), which appears, in one
form or another, in all event generators. For instance,
the H1 found \cite{H1} significant disagreement
between their data on jets with $E_T^{\mathrm jet}\approx 7$ GeV
and PYTHIA and HERWIG event generators
\underline{without} the sue. For PYTHIA this discrepancy
disappeared when the multiple interaction option with a
very small partonic $p_T^{\mathrm min}= 1.2$ GeV was used.
The presence of sue may also be the
cause of the discrepancy between NLO calculations of
dijet cross--sections in photoproduction and ZEUS data
\cite{ZEUS,Au} in the region
$x_{\gamma}\le 0.75, E_T^{\mathrm jet}\le 10$ GeV.

In HERWIG the sue
is modelled as the soft interaction of two beam remnant clusters,
which accompanies the basic hard parton--parton scattering. The
main effect of this soft collision is the redistribution of the
(usually large) longitudinal
energy of the two beam remnant clusters. Without sue
these remnant clusters decay into just
a few particles, separated by a large rapidity gap.
In a sue event this longitudinal energy is redistributed among many
softer particles which populate central region in pseudorapidity
and thus provide ``pedestal'' under the jets.
The strength and frequency of the sue
is governed by two parameters, which can be tuned to data.
As there are so far no conclusive results on the magnitude of sue
needed by the data
we performed our studies for two extreme options: no sue and sue
in each event and with all the energy of beam remnants used up
for the soft collision.
The differences between these two scenarios, shown in Figs. 2,3,
illustrate the importance of a good quantitative understanding of
nonjet physics for the determination of jet properties.

We defined jets using the CDF cone algorithm with $R=1$
and required $E_T\ge 5$ GeV in $\gamma$p CMS.
The crucial question is how much transverse energy of the sue
populates the region in $\eta$ where jets are found. For a given
$x_{\gamma} \equiv
(E_T^{(1)}\exp(-\eta_1)+E_T^{(2)} \exp(-\eta_2))/2E_{\gamma}$, 
$x_p, W$ and $p_T^{\mathrm jet}$ simple
kinematics allows us to determine the intervals of $\eta^{\mathrm jet}$
and $\eta^{\mathrm remn}$, populated by jets and by soft
particles from beam remnant fragmentation respectively.
In Fig. 1b they correspond, as a function of a $x_{\gamma}$ and for
fixed typical $x_p=0.05, p_T^{\mathrm jet}=5$ GeV and two values of
$W=75,200$ GeV,
to intervals on the $y$ axis between
the solid (for jets) and dashed curves (for beam remnants).
We see that except for $x_{\gamma}$ very close to 1,
jets can bath in a pool of soft particles from beam fragmentation.
Whether they indeed do so depends on the way beam remnants fragment.
In Fig. 2 we show, for events with and without sue, normalized
legoplots of $E_T$ as a function of $\eta^{\mathrm jet}$
and $x_{\gamma}$ for jets and $\eta^{\mathrm remn}$ and $x_{\gamma}$
for particles from beam remnants.
Without the sue jets and beam remnants
overlap significantly only for large $x_{\gamma}$,
while for events with the sue they do so for all values of
$x_{\gamma}$. Moreover,
the amount of transverse energy under the jets coming from beam
remnant
fragmentation is almost negligible for events without the sue, but
becomes quite sizeable for the sue option.
\begin{figure}[t]
\begin{center}
\epsfig{file=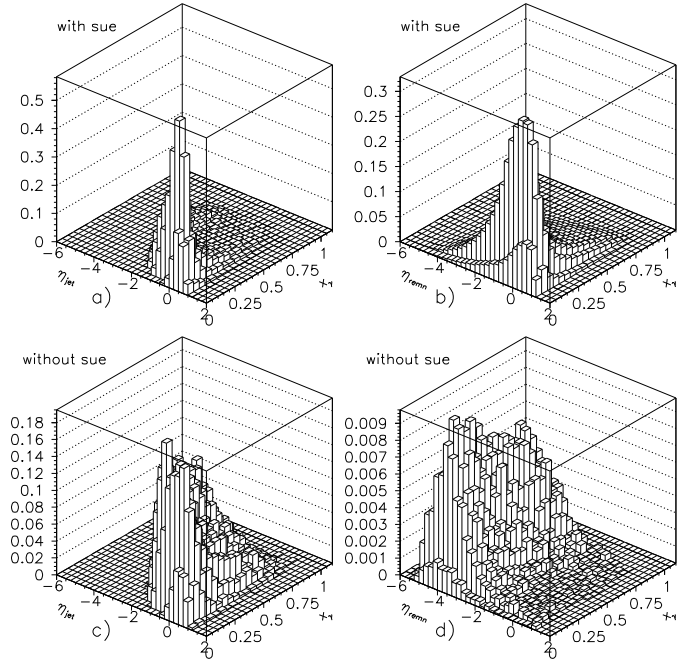,height=9cm}
\end{center}
\caption{Transverse energy flow per event as a function of
$x_{\gamma}$ and pseudorapidity of jets or particles from beam remnants
in events with (a,b) and without (c,d) the sue.}
\end{figure}
There is always some transverse energy under the jets which distorts its
properties, but above certain value it makes the theoretical predictions
unstable and thus unreliable. Too much transverse energy from sue leads
to fake hard jets, which significantly increase the corresponding
cross--section. We studied this effect in HERWIG by generating events
with $p_T^{\mathrm part}$ of the basic partonic
subprocess starting from 2 GeV and looked for jets with
$E_T$ between 5 and 12 GeV. The results, plotted in Fig. 3a, show
that without the sue (dashed curves), the contributions to jets with a
given $E_T^{\mathrm jet}$ come mainly from partons with $p_T^{\mathrm
part}\doteq E_T^{\mathrm jet}+1$ and drop off rapidly below and more
slowly above $E_T^{\mathrm jet}$. Summing the contributions of all
events with $p_T^{\mathrm part}>2$ GeV yields the dashed curves in Fig.
3b. The jet cross--sections are reasonably stable with respect to
the variation of the minimal partonic $p_T^{\mathrm min}$ down to
$E_T^{\mathrm jet}=5$ GeV. For events with sue the situation, described
by solid curves in Fig. 3, is substantially different. Even events  with
$p_T=2$ GeV can fake jets with $E_T^{\mathrm jet}$ below 10 GeV and this
probability in fact increases with decreasing $p_T^{\mathrm part}$!
A closer scrutiny shows that these fake hard jets are characterized by
almost no correlation between the momenta of basic hard parton
scattering process and the observed jets. Only
for $E_T^{\mathrm jet}$ above roughly 10 GeV we get the situation
analogous to that of dashed curves. The integrated jet cross--sections
in Fig. 3b are then highly unstable with respect to the minimal partonic
$p_T^{\mathrm min}$, which makes the theoretical predictions
unreliable. In order to make sensible comparisons between theory and
data the amount of transverse energy under the jets (jet pedestals)
must be determined experimentally and subtracted from measured jets
transverse energy. This is in particular true for comparisons at the
NLO, where theoretical calculations are available on partonic level
only. Comparing theoretical prediction directly with observed jets makes
sense only if the theory describes quantitatively also the transverse
energy flow \underline{outside} the jets.
This point will be discussed in more detail in \cite{my2}.
\begin{figure}
\begin{center}
\epsfig{file=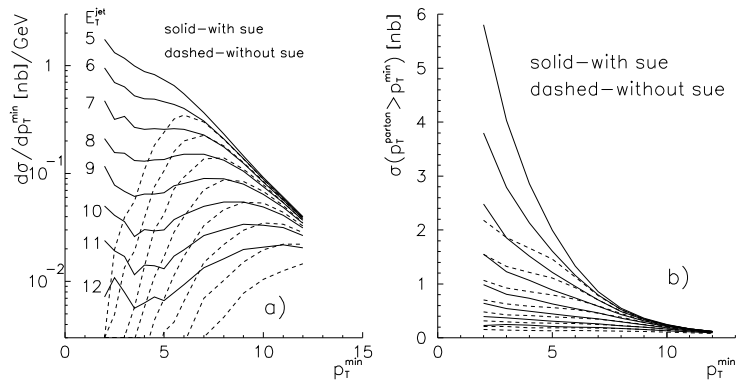,height=5cm}
\end{center}
\caption{The contribution of a given events with given $p_T^{\mathrm
part}$ to the cross--section for production of jets with $E_T^{\mathrm
jet}$ between 5 and 12 GeV (a) and sums of these contribution above
a given minimal partonic $p_T^{\mathrm min}$ (b).}
\end{figure}

\vspace*{0.2cm}
%\section*{Acknowledgments}
\noindent
This work has been supported by the Grant Agency of
the AS of the Czech Republic under the grants
No. A1010602 and A1010619.

\end{document}